\documentclass[12pt]{article}
\usepackage{epsfig,float,amssymb,stmaryrd,latexsym}
\DeclareMathSymbol{\varGamma}{\mathord}{letters}{"00}

\newcommand{\be}{\begin{equation}}
\newcommand{\ee}{\end{equation}}

\newcommand{\ds}{\displaystyle}
\newcommand{\vp}{\varphi}
\newcommand{\vt}{{\vec \tau}}
\newcommand{\vk}{{\vec k}}

\title{\hfill{\small FZJ--IKP(TH)--2007--04} \\[1.8em]
\bf Two--photon decays of hadronic molecules}

\author{C. Hanhart$^1$, Yu. S. Kalashnikova$^{2}$, A. E. Kudryavtsev$^{2}$,
and A. V. Nefediev$^{2}$}

\date{\small\em $^1$Institut f\"{u}r Kernphysik, Forschungszentrum J\"{u}lich GmbH, D--52425 J\"{u}lich, Germany \\[2mm]
$^2$Institute of Theoretical and Experimental Physics, 117218, B. Cheremushkinskaya 25, Moscow, Russia\\}

\begin{document}
\maketitle

\begin{abstract}
In many calculations of the two--photon decay of hadronic molecules,
the decay matrix element is estimated using the wave function at the
origin prescription, in analogy to the two--photon decay of
parapositronium. We question the applicability of this procedure to
the two--photon decay of hadronic molecules for it introduces an
uncontrolled model dependence into the calculation. As an alternative
approach, we propose an explicit evaluation of the hadron loop. For
shallow bound states, this can be done as an expansion in powers of
the range of the molecule binding force $1/\beta$. In the leading
order one gets the well-known point-like limit answer. We estimate, in
a self--consistent and gauge invariant way, the leading range
corrections for the two--photon decay width of weakly bound hadronic
molecules emerging from kaon loops.  We find them to be small, of
order ${\cal O}(m\varepsilon/\beta^2)$, where $m$ and $\varepsilon$
denote the mass of the constituents and the binding energy,
respectively.  The role of possible short--ranged operators and of the
width of the scalars remains to be investigated.
\noindent
\end{abstract}

\pagebreak

\section{Introduction}

Hadronic molecules are bound states of two hadrons held together
by the strong interaction --- clearly to be distinguished from
the so-called hadronic atoms, where the two hadrons are bound by
the Coulomb interaction. In the latter case the strong interaction
only leads to a slight shift in the binding energies (and
an additional width). Hadronic atoms can nowadays be produced
in laboratories almost routinely. Hadronic molecules, on the other
hand, might well be part of the hadron spectrum but are not yet
identified unambiguously. In recent years evidence has grown that
a few of the large number of known scalar mesons might be of
molecular character. For recent reviews on
the meson spectrum, with emphasis on the
heavy states, see Refs.~\cite{klempt,swanson,rosner}.

It was argued for many years that the studies of the two--photon decay
of scalars could distinguish among different scenarios for scalar
meson structure. 
One of the most studied cases is that of the light scalar
mesons $a_0(980)$ and $f_0(980)$ and
indeed, the predictions of various models for these
 differ drastically.  Assuming them to be $q
\bar q$ states made of light quarks, one gets about $1.3$ to $1.8$ keV
for the $f_0(980)\to\gamma\gamma$ width in the relativistic quark
model \cite{Munz}, while, under the $s \bar s$ assumption, the
two--photon width of the $f_0(980)$ is calculated to be about $0.3\div
0.5$ keV \cite{Del}.  Within the molecular model for scalars, the
predictions vary from $0.2$ keV in Ref.~\cite{OO}\footnote{In this
paper the chiral unitary approach is used. That
the scalars produced are to be interpreted as
dynamically generated is shown in Ref.~\cite{jose}.
For a somewhat different view on this subject see Ref.~\cite{jaffe}.} to $0.6$ keV in
Ref.~\cite{Barnes1} and to $6$ keV in Ref.~\cite{SK}.  In the present paper we
demonstrate that the technique used in Refs. \cite{Barnes1,SK} has a
large theoretical uncertainty. We also show that a gauge--invariant
treatment of the two--photon decay amplitude of the $K \bar K$
molecule yields the value of the $\gamma\gamma$ width for the
$f_0(980)$ close to $0.2$~keV.

It is well known since long ago that the two--photon decay rate for
the parapositronium is very well approximated by the product of the
square of the wave function at the origin times the $e^+ e^-$
annihilation rate at rest \cite{Landau,pos1}.  This was taken as a
recipe by many authors and was applied also to calculate the
two--photon decay rates of hadronic molecules \cite{Barnes1,SK}. In
this paper we argue that this procedure leads to wrong
results. Instead we propose to calculate explicitly the hadron loops
employing an expansion in the range of forces, $1/\beta$. Then the
leading term assumes a point-like molecule
vertex and the two--photon decay of a scalar meson is found to be
\begin{equation}
\Gamma_{\gamma\gamma}=\zeta^2\left(\frac{\alpha}{\pi}\right)^2\sqrt{m\varepsilon}\left(\frac{2m}{m_S}\right)
\left[\left(\frac{2m}{m_S}\right)^2\arcsin^2\left(\vphantom{\frac{2m}{m_S}}\frac{m_S}{2m}\right)-1\right]^2,\quad m_S=2m-\varepsilon,
\label{beauty}
\end{equation}
where $m_S$ is the scalar meson mass, $\varepsilon$ denotes the
binding energy of the hadronic molecule and $m$ the mass of the
constituents --- for simplicity we only study systems with
constituents of equal mass --- and $\alpha=e^2/4\pi$ denotes the
fine--structure constant.  The factor $\zeta$ is different from 1, if
not all constituents participate in the decay. For example, in case of
the $f_0$ only the charged kaons couple to the photons in leading
order and therefore $\zeta^2=1/2$.  We also calculate the leading
range corrections to the two--photon decay rate. They turned out to be
suppressed by a factor $m\varepsilon/\beta^2\ll 1$, since we focus on
shallow bound states.

The paper is organized as follows: in the next section we present some
very general arguments, why the wave function at the origin
cannot be used to calculate the decay of hadronic molecules. This will
be demonstrated explicitly in the subsequent sections: in
Section~\ref{bsa} we give the general formulae for the two--photon
decay of bound states that allow us to investigate two limits: the
weak coupling limit --- that leads to the wave function at the origin
prescription --- is discussed in chapter~\ref{secwcl} and the limit of
the point-like interactions in chapter~\ref{seczrl}. In
Section~\ref{seclrc} the leading range corrections to the latter limit
are calculated. We close with a summary and outlook.

\section{The relevant scales}

Before we go into details let us present some general arguments
why
the wave function at the origin should not be used to calculate the 
two--photon decay of hadronic
molecules. The most obvious argument is that we simply do not
know the
wave function at the origin. In contrast to the parapositronium decay, the
equations solved for hadronic molecules are not solved using the
fundamental degrees of freedom. Instead one typically works with
conveniently chosen interpolating fields --- and this choice influences
the short--range behavior of the molecule wave function. For the deuteron wave
function this is to some extend discussed in
Ref.~\cite{mitandreas}. Only the tail of the wave function is
completely determined by the binding energy and is therefore known
model independently. Our ignorance about the wave function at the
origin translates into a large spread for predictions for the
corresponding two--photon decay rate of, say, the light scalar
mesons, from $0.6$ keV in Ref.~\cite{Barnes1} to $6$ keV in ref.~\cite{SK}\footnote{Also
the recent attempt to improve on the wave function at the 
origin formula presented in Ref. \cite{lemmernew} is not
a solution, for it suffers from the same ignorance and, in addition,
 leads to a violation of gauge invariance, as
explained below.}.

\begin{figure}[t]
\begin{center}
\epsfig{file=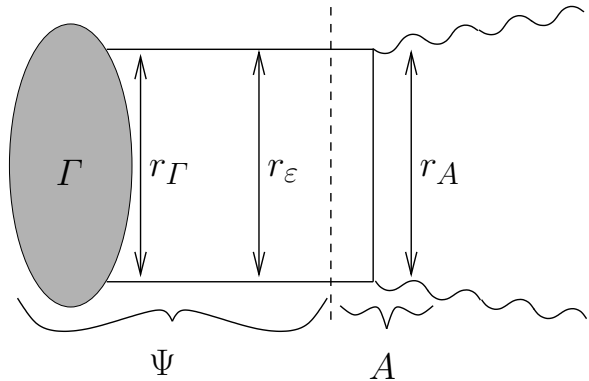,height=5cm}
\caption{Illustration of the different parts relevant for the decay of a hadronic molecule into two photons.
The vertex for the transition of the molecule into its constituents is denoted by $\varGamma$, the corresponding
wave function by $\Psi$, and the annihilation potential by $A$. Solid and wavy lines denote the propagation of the constituents of the
molecule and of the photons, respectively.}
\label{scales}
\end{center}
\end{figure}

The second argument is that any transition matrix element using the
wave function at the origin meets certain problems with gauge invariance. In
case of positronium this is a minor effect, since the violations
are suppressed by at least one extra power in the fine structure constant
$\alpha$. In case of hadronic molecules, however, this violation might
well be more severe. This will be discussed in some detail below.

The third argument is that the hierarchy
of scales in case of the decay of hadronic molecules is very
different
to that of positronium decay. The individual parts of the
decay
are illustrated in Fig.~\ref{scales}. First of all there is the
molecule
vertex $\varGamma$ for the decay of the molecule into its constituents
--- here two mesons\footnote{For simplicity we talk of mesons only for
the constituents. Note that the reasoning does not need to be
changed in the presence of fermions.}.
Next come the two meson propagators. The final piece is the
annihilation potential $A$, given
by the photon--meson vertices and the intermediate meson propagator.
Corresponding to the building blocks there are three scales
relevant for the two--photon decay of the bound state.
To begin with, there is the intrinsic scale $r_\varGamma$ of
the vertex function $\varGamma$ set
by the dynamics of the bound--state formation.
An additional scale $r_\varepsilon\sim 1/\kappa$ appears due to
presence of the bound state,
where we defined the binding momentum
$\kappa=\sqrt{m\varepsilon}$,
$m$ is the mass of the molecular constituents.
The third scale is given by the range of the annihilation
potential. For a shallow bound state, the energy carried away by the individual photons is of the order of $m$.
Consequently, the range of the annihilation is given by $r_A=1/m$.

Let us consider parapositronium decay from the point of view
of the hierarchy of the scales introduced above. We clearly deal
with a nonrelativistic
system with the binding energy $\varepsilon=\alpha^2
m_e/4$, with $m_e$ denoting the electron mass.
Note, it is the parameter
$\kappa=\sqrt{m_e\varepsilon}=\alpha m_e/2$
that defines the long--range piece of the molecular wave function, which takes
the form $\Psi(r)=\sqrt{\kappa^3/\pi}\exp{(-\kappa r)}$.
The vertex function depends only on the electron three--momentum $\vec{p}$ and is trivially related to the bound--state wave function \cite{moldec}:
\be
\varGamma(\vec{p})=\sqrt{2m_S}(\vec{p}\,^2+\kappa^2)\psi(\vec{p}),
\label{GPsi}
\ee
with $m_S$ being the positronium mass. We denote wave functions
in coordinate space by $\Psi$ and their Fourier--transforms in 
momentum space by $\psi$. An explicit calculation
with the positronium wave function yields
\be
\varGamma(\vec{p}) \propto \frac{1}{\vec{p}\,^2+\kappa^2},
\ee
so that
$r_\varGamma \sim 1/\kappa$ in the positronium case.
Finally, as discussed above, we have $r_A\sim 1/m_e$. Therefore there is the hierarchy of scales
\begin{equation}
\mbox{Case A:} \qquad r_A \ll r_\varepsilon \approx
r_\varGamma.
\label{scpos}
\end{equation}
Thus, in case of the decay of positronium, the annihilation process is well approximated as taking place at the origin and
consequently the decay amplitude scales to an excellent approximation with the wave function at the origin.

Quite an opposite situation takes place for molecular
hadronic systems. Indeed, in this case the scale of the
vertex function is defined by the range of binding forces
$1/\beta$. If one deals with a loosely bound state formed
by zero--radius forces ($\beta\to\infty$) the hierarchy is
\begin{equation}
\mbox{Case B:} \qquad r_\varGamma \ll r_A \ll r_\varepsilon.
\label{schad}
\end{equation}
Then annihilation process cannot be described with the wave
function at the origin prescription.

To see which case (case A or case B) is more adequate for hadronic
molecules let us focus on the two--photon decay of the $f_0(980)$ as a
kaon molecule. Then we have $\beta \sim m_\rho$, where $m_\rho$ is the
mass of the $\rho$-meson, the lightest meson participating in the
meson exchange between kaons (there is no one--pion exchange between two pseudoscalars),
$\varepsilon < 0.1 m$, and, again, $r_A \sim 1/m$. This leads to
\begin{equation}
r_\varGamma < r_A \ll r_\varepsilon. 
\end{equation}
Comparing this to
Eqs.~(\ref{scpos}) clearly shows that the decay of hadronic molecules
calls for a very different treatment as compared to that for the decay
of positronium. The question arises if it is at all possible to give a
simple recipe to calculate such a two--photon decay of, say, the
$f_0$. In this paper we argue that the corresponding decay amplitude
is well approximated by a kaon--loop integral evaluated in the limit
of a point-like decay vertex ($\beta\to\infty$). We thus propose to
work in the limit (\ref{schad}), as the zeroth approximation, and
build finite--range corrections in powers of $1/\beta$ to the leading
term.  Naively one would expect such corrections of order of
$(m/\beta)^2$ which, in case of the $f_0$, turn out to be of the order
of 40\%. However, an explicit calculation presented below shows that
the leading range corrections in $1/\beta$ only scale as $(\kappa/\beta)^2$
which, in case of the $f_0$, turns out to be of the order of 1\%.
Therefore, in case of the $f_0$ the corrections to
the point--like formula, Eq.~(\ref{beauty}), are expected to be at most
of order $(m/\beta)^4\sim 15$\%.

\section{Bethe--Salpeter approach}
\label{bsa}

\begin{figure}[t]
\begin{center}
\epsfig{file=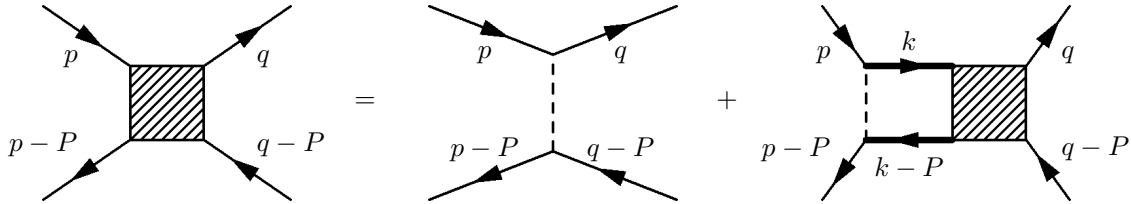,width=15cm}
\caption{Graphical representation of Eq.~(\ref{Tk}) for the full meson--meson scattering amplitude.}\label{Te}
\end{center}
\end{figure}

In this section we employ an explicitly gauge invariant approach based
on the Bethe--Salpeter equation for the molecule vertex in order to
illustrate in more detail the interplay of the various scales. For
scattering amplitudes similar formalisms were discussed in
Refs.~\cite{sasha,bugra,haberzettel}.  The relevant equations in the
two limiting cases A and B introduced above will appear as special
cases of these general equations.

Consider a Lorentz--covariant theory describing the meson--meson interaction via
a potential $V(p-k)$ which possesses the inverse interaction range $\beta$.
A priori no assumption needs to be made on the structure of $V$, however, to
keep the expressions simple in this very general discussion we assume that
there is no charge flow in the potential.
As a consequence there will be no
meson exchange currents, when we include photons. This situation is 
naturally realized for potentials given by $t$--channel exchanges of neutral particles.
This gives rise to the so--called ladder approximation for the scattering
equation that we will refer to in the following for simplicity.

In practice the just described restriction on the potential implies
the omission of many diagrams without a priori
justification. However, the goal of this section is to demonstrate
that in gauge invariant approaches self--energies get linked to
scattering potentials. We will not draw any quantitative conclusions
from the considerations in this section. In contrast to this, when we
discuss the leading finite range corrections in Sec. \ref{seclrc} as
well as Appendix \ref{excur} we do not need to make any additional assumptions
and --- to this order in the range of forces --- the problem is
solved exactly.
next section the effect of charge exchange is explained in Appendix \ref{excur}.

\begin{figure}[t]
\begin{center}
\epsfig{file=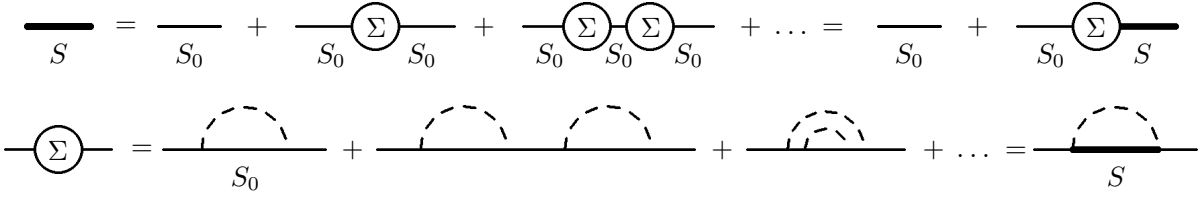,width=16cm} \caption{Graphical
representation of the Dyson equations for the dressed meson propagator
and for the meson self--energy, Eq.~(\ref{Gk}).}\label{DE}
\end{center}
\end{figure}

Scattering of two mesons can be described by the equation (see Fig.~\ref{Te})
\be
T(p,q,P)=V(p-q)-i\int\frac{d^4k}{(2\pi)^4}S(k)S(k-P)V(p-k)T(k,q,P),
\label{Tk}
\ee
where $P_\mu$ is the total momentum of the bound state and $p_\mu$ is the four--momentum of one of its constituents.
The propagators given are solutions of the Dyson equation, presented in
the graphical form in Fig.~\ref{DE} --- if we again assume $V$ to refer to the emission and absorption
of a neutral meson, we here work in the rainbow approximation
\be
S^{-1}(p)=S_0^{-1}(p)-\Sigma(p),\quad S_0(p)=\frac{1}{p^2-m_0^2},\quad\Sigma(p)=-i\int\frac{d^4k}{(2\pi)^4}S(k)V(p-k),
\label{Gk}
\ee
with $m_0$ being the bare meson mass. The physical meson mass $m$ appears as the pole of the dressed propagator $S(p)$.

If there exists a bound--state with the mass $m_s$, we may
define the
corresponding vertex function $\varGamma(p,P)$ as
the solution of a homogeneous Bethe--Salpeter equation
\be
\varGamma(p,P)=-i\int\frac{d^4k}{(2\pi)^4}S(k)S(k-P)V(p-k)\varGamma(k,P),
\label{BS}
\ee
which is to be evaluated at $P^2=m_s^2$. The bound--state
vertex is normalized through the condition \cite{LS}
\be
-i\int\frac{d^4k}{(2\pi)^4}\varGamma^2(k,P)\frac{\partial}{\partial P_\mu}S(k)S(k-P)=2P_\mu,
\label{mnc}
\ee
which relates the vertex $\varGamma(p,P)$ to the bound--state mass.

\begin{figure}[t]
\begin{center}
\epsfig{file=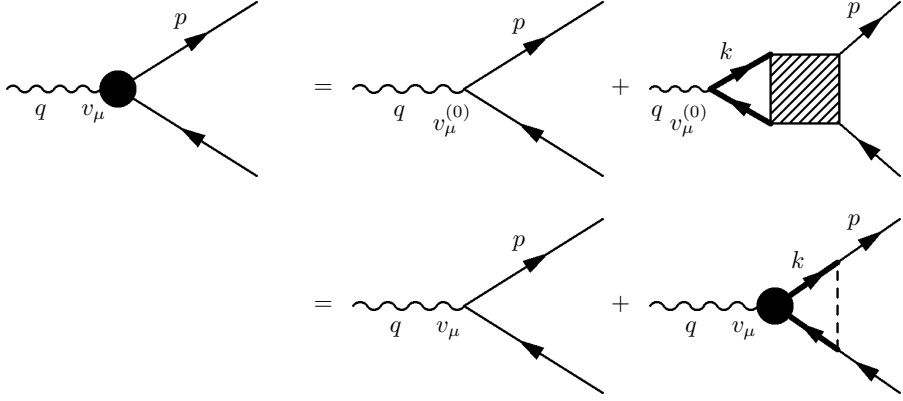,width=12cm}
\caption{Graphical representation of Eq.~(\ref{vdr}) for the dressed photon emission vertex.}\label{vdr0}
\end{center}
\end{figure}

To describe radiative processes one should first define the
dressed photon emission vertex for a meson.
In the absence of charge flow in the potential $V$ this is (see Fig.~\ref{vdr0})
\begin{eqnarray}
v_\mu(p,q)&=&v^{(0)}_\mu(p,q)-i\int\frac{d^4k}{(2\pi)^4}T(p,k,q)S(k)S(k-q)v_\mu^{(0)}(k,q) \nonumber\\[-5mm]
\label{vdr}\\
&=&v^{(0)}_\mu(p,q)-i\int\frac{d^4k}{(2\pi)^4}V(p-k)S(k)S(k-q)v_\mu(k,q)\nonumber,
\end{eqnarray}
where \be v^{(0)}_\mu(p,q)=(2p-q)_\mu \ee and $q_\mu$ and $p_\mu$ are
the emitted photon and the emitting meson momenta,
respectively. As follows
from Eqs.~(\ref{Tk}) and (\ref{vdr}), the dressed vertex $v_\mu(p,q)$
obeys the Ward identity, \be q_\mu v^\mu(p,q)=S^{-1}(p)-S^{-1}(p-q).
\label{Wi}
\ee

\begin{figure}[t]
\begin{center}
\begin{tabular}{ccc}
\epsfig{file=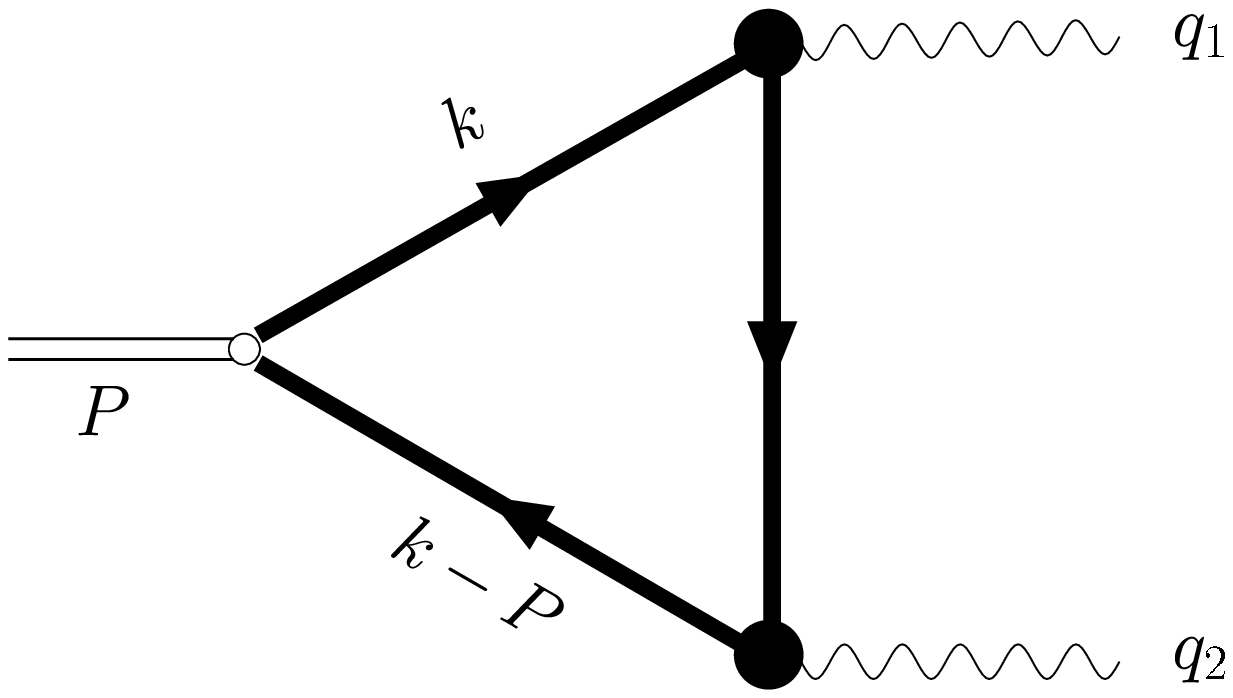,width=5cm}&\epsfig{file=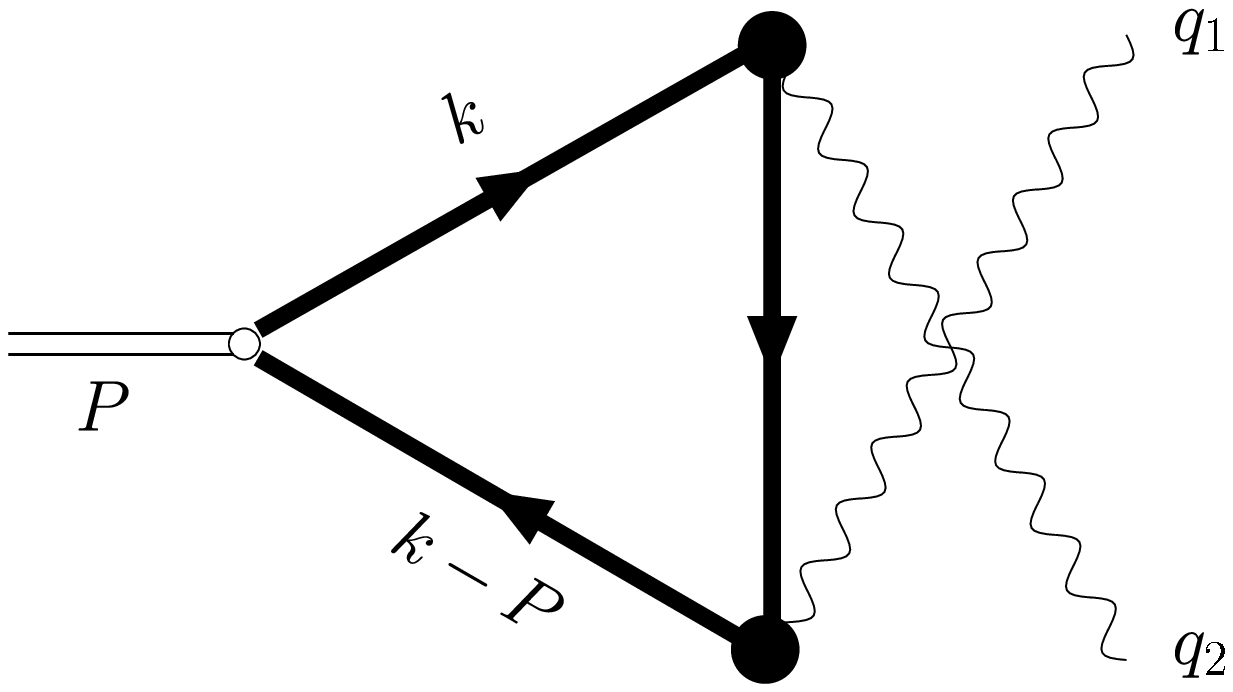,width=5cm}&\raisebox{3mm}{\epsfig{file=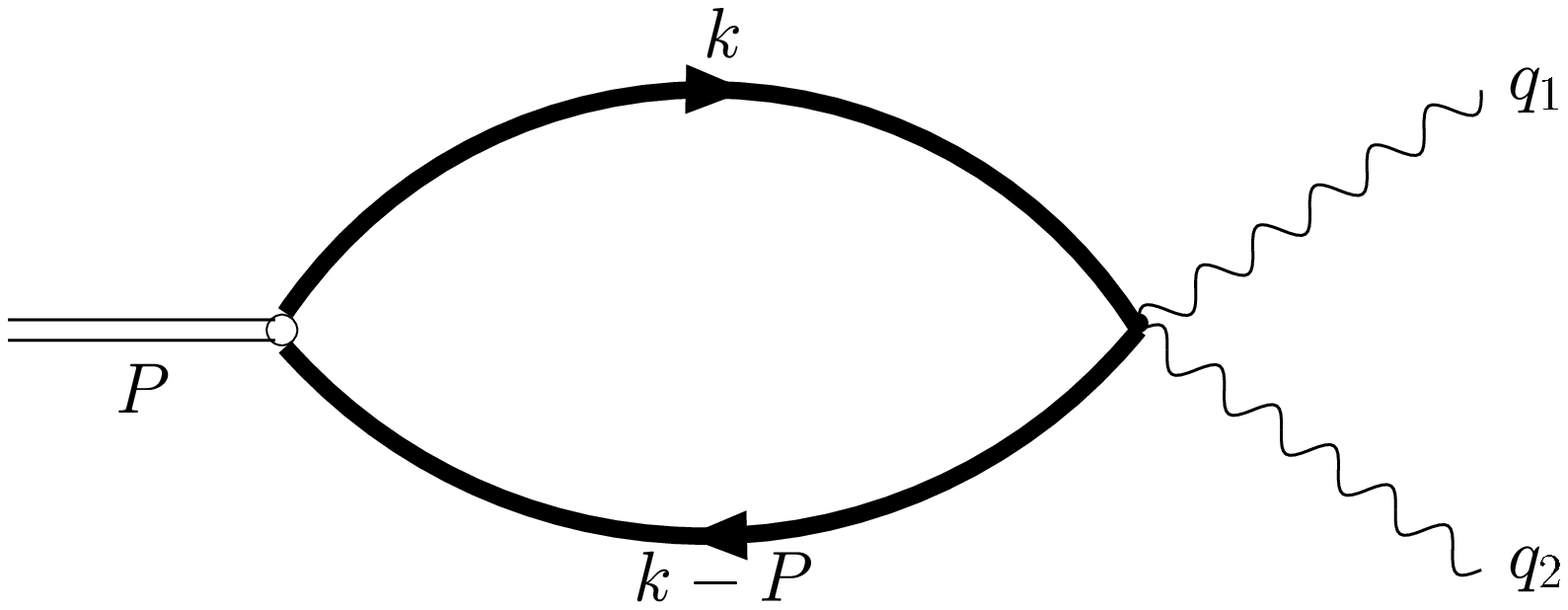,width=6cm}}\\
(a)&(b)&(c)\\
\end{tabular}
\end{center}
\caption{Diagrams contributing to the scalar decay amplitude.}\label{3d}
\end{figure}

The two--photon decay amplitude for the bound state can now be evaluated with
the help of the diagrams depicted in Fig.~\ref{3d} and with the
dressed vertices and propagators involved (notice that the seagull vertex in
Fig.~\ref{3d}(c) need not be dressed since, due to the Bethe--Salpeter Eq.~(\ref{BS}), the corresponding diagrams are
already included into the definition of the scalar vertex). The resulting
transition matrix element is
\be
W=\varepsilon_{1\mu}^*\varepsilon_{2\nu}^*W^{\mu\nu},
\ee
where
\begin{eqnarray}
W_{\mu\nu}&=&e^2\int\frac{d^4k}{(2\pi)^4}\varGamma(k,P){v_{\mu}(k,q_1)v_{\nu}(k-q_1,q_2)}S(k-q_1)S(k-P)S(k)\nonumber\\
&+&e^2\int\frac{d^4k}{(2\pi)^4}\varGamma(k,P){v_{\nu}(k,q_2)v_{\mu}(k-q_2,q_1)}S(k-q_2)S(k-P)S(k)\label{WG}\\
&-&2g_{\mu\nu}e^2\int\frac{d^4k}{(2\pi)^4}\varGamma(k,P)S(k-P)
S(k)\nonumber,
\end{eqnarray}
with $q_{1,2}$ and $\epsilon_{1,2}^*$ being the
four--momenta
and the
polarization vectors of the two photons.
The quantity $W^{\mu\nu}$ appears to be gauge invariant,
\be
W^{\mu\nu}q_{1\mu}=W^{\mu\nu}q_{2\nu}=0.
\label{chcon}
\ee
To show this one may use the Ward identity, Eq.~(\ref{Wi}), to write
\begin{eqnarray}
\nonumber
W_{\mu\nu}q_1^{\mu}&=&e^2\int\frac{d^4k}{(2\pi)^4}\varGamma(k,P)S(k-P)S(k-q_1)v_{\nu}(k-q_1,q_2)\\
\nonumber
&-&e^2\int\frac{d^4k}{(2\pi)^4}\varGamma(k,P)S(k)S(k-P)v_{\nu}(k-q_1,q_2)\\
&+&e^2\int\frac{d^4k}{(2\pi)^4}\varGamma(k,P)S(k)S(k-P)v_{\nu}(k,q_2)\label{qW}\\
\nonumber
&-&e^2\int\frac{d^4k}{(2\pi)^4}\varGamma(k,P)S(k)S(k-q_2)v_{\nu}(k,q_2)\\
&-&2q_{1\nu}\int\frac{d^4k}{(2\pi)^4}\varGamma(k,P)S(k)S(k-P)\nonumber.
\end{eqnarray}
Now, using the bound--state equation for $\varGamma(k,P)$ and the second line of Eqs.~(\ref{vdr}) one may write
\begin{eqnarray}
&&\int\frac{d^4k}{(2\pi)^4}\varGamma(k,P)S(k-P)S(k-q_1)v_{\nu}(k-q_1,q_2)\nonumber \\[-5mm]
\\
&& \qquad \qquad =\int\frac{d^4l}{(2\pi)^4}\varGamma(l,P)S(l-P)S(l)\left(v_{\nu}(l-q_1,q_2)
-v_{\nu}^{(0)}(l-q_1,q_2)\right).\nonumber
\end{eqnarray}
The same manipulations applied to the fourth line of Eq.~(\ref{qW}) lead to Eq.~(\ref{chcon}).

We therefore see that it is necessary that the vertex function $\varGamma$ and the photon--meson vertices are
constructed consistently in order to get gauge invariant amplitudes. In other words, using in the
expression for the decay amplitude the molecule wave function together with bare vertices and propagators,
inevitably leads to the violation of gauge invariance.

For the decay $S\to\gamma\gamma$ involving real photons,
Eqs.~(\ref{chcon}) imply that
\begin{equation}
iW=M(P^2)[q_1^{\nu}q_2^{\mu}-g^{\mu\nu}(q_1q_2)]
\epsilon_{1\mu}^*\epsilon_{2\nu}^*,\quad P=q_1+q_2.
\label{vertex}
\end{equation}
Then, for the scalar of the mass
$m_S$, the total width of such a decay can be evaluated as
\begin{equation}
\Gamma_{\gamma \gamma}=\frac{m_S^3}{64\pi}|M(m_S^2)|^2,
\label{observ}
\end{equation}
where the identity of the photons in the final state is
taken into account in
the overall coefficient in Eq.~(\ref{observ}).

Equation~(\ref{WG}) is still general (up to the absence of exchange currents) and we may study it in both limits: case A
(Eq.~(\ref{scpos})) as well as case B (Eq.~(\ref{schad})).
Note, in order to simplify Eq.~(\ref{WG}) we need to assume the coupling to be weak. 
In a situation, where case A holds for strong couplings, the full system of coupled equations
needs to be solved.
 In the next subsections both limits are discussed individually.

\subsection{Case A in the weak coupling limit}
\label{secwcl}

In case A, $r_\varGamma\gg r_A$.
As we shall see, the decay width in this limit can be derived
from the general expression of Eq.~(\ref{WG}) under the assumption of
weak coupling. Then one may neglect the dressing effects and
self--energies altogether --- in Eq.~(\ref{WG}) all propagators
and vertices can be replaced by the bare ones.
As outlined above, this necessarily implies a certain violation
of gauge invariance, however, those
violations are suppressed by at least one power in the coupling
that is assumed to be small. In the limit considered, the typical momentum
in the loop is very small and one may replace $k_0$  in the photon vertices as well as in the
strong vertex $\varGamma$ by $m$. Then one may write in the
rest frame of the scalar ($P^\mu=(2m-\varepsilon,\vec{0})$):
\begin{eqnarray}
-i\int\frac{dk_0}{2\pi}S(k)S(k-P)&\approx&\frac{1}{4m}\frac{1}{\vec{k}\,^2+\kappa^2},\nonumber\\
\\[-3mm]
-i\int\frac{dk_0}{2\pi}S(k-q_1)S(k-P)S(k)&\approx& \left.S(k-q_1)\right|_{k^0=q_1^0}\left(-i\int\frac{dk_0}{2\pi}S(k)S(k-P)\right),\nonumber
\end{eqnarray}
where only the leading pole is kept and non--relativistic
kinematics is used for the mesons with momentum $\vec{k}$.
A similar expression
appears for $q_1\leftrightarrow q_2$. What remains to be
evaluated now is the three--dimensional integral
\be
W_{\mu\nu}=-\zeta \, e^2\sqrt{\frac{2}{m}}\int\frac{d^3k}
{(2\pi)^3}\psi(\vec{k})
\left(\frac{(2\tilde k-q_1)_\mu (2\tilde k-P-q_1)_\nu}
{m^2+(\vec k-\vec q_1)^2}+
(\mu\leftrightarrow\nu,q_1\leftrightarrow q_2)+2g_{\mu
\nu}\right),
\label{nonrel}
\ee
where we used the relation (\ref{GPsi}) and defined
$\tilde k^\mu=(m,\vec k)$.
The term in parenthesis refers to
the annihilation potential.
By assumption we have $r_A\ll r_\varGamma$ which translates
into $\vec{k}\ll\vec{q}_{1,2}$. Under this condition one may
neglect all
$\vec{k}$ dependence in this term, which then reduces to the
annihilation potential at rest, and pull it out of the integral.
The remaining integral is nothing but the definition of the
wave function at the origin (in coordinate space). These altogether yield gauge
invariant answer (\ref{vertex}) for the amplitude, with
\be
M(P^2)=i \zeta e^2\frac{\Psi(0)}{m^{5/2}}.
\ee
Thus one arrives at the following expression for the two--photon
decay width for the limiting case~A,
\be
\Gamma_{\gamma\gamma}^A=2\zeta^2\frac{\pi\alpha^2}{m^2}|\Psi (0)|^2.
\label{GammaKK}
\ee
Note that the final answer is gauge invariant, but this is
true only for the leading term in an expansion in the potential $V$
for the transition matrix elements. As explained, if one wants to 
improve the accuracy of Eq.~(\ref{GammaKK}) it is insufficient
to just keep the momentum dependence in Eq.~(\ref{nonrel}), as proposed
in Ref.~\cite{lemmernew}, but also the meson self--energies are
to be kept explictly.
Consequently, Eq.~(\ref{GammaKK}) should only be applied in the
weak coupling limit.

There exists a prescription to calculate the
two--photon decay
amplitude by contracting the on-shell decay amplitude with
the bound--state wave function $\psi(\vk)$ (see, e.g.
\cite{BCL}):
\be
W\propto \int\frac{d^3k}{(2\pi)^3}\psi(\vk)
\left[W(K^+(\vk)K^-(-\vk)\to\gamma\gamma)\right].
\label{M1}
\ee
Since gauge invariance is preserved for the on-shell
amplitude $W(K^+(\vk)K^-(-\vk)\to\gamma\gamma)$,
then the full amplitude (\ref{M1}) proves to be gauge
invariant automatically.
In the leading nonrelativistic approximation the $\vec k$ dependence of $W$ can be neglected, so that 
Eq.~(\ref{M1}) is identical to Eq.~(\ref{GammaKK}). However, in general  Eq.~(\ref{M1})
violates energy
conservation: in the c.m. frame the use of the on-shell
amplitude in Eq.~(\ref{M1}) implies that the kaon energies
$k_{10}=k_{20}=\sqrt{\vk^2+m^2}$, while energy
conservation requires $k_{10}=k_{20}=m_S/2$. This problem
is discussed in detail in Ref.~\cite{Pestieau}.

Simple recipes to restore gauge invariance in the presence of non--trivial
vertex functions through new contact diagrams with the derivatives of
this vertex, successfully used for decays like $\phi\to \gamma f_0$~\cite{moldec}, fail, since the photons are
not soft. As a result, gauge invariance, preserved in the point-like limit, appears
broken already to order $(m/\beta)^2\sim 0.4$  (see Appendix \ref{fail} for the details), where we used for illustration
with $m=m_K$ and $\beta=m_\rho$ the parameters
relevant for the $f_0$. In the previous section we showed that the inclusion
of the scalar vertex structure in a gauge
invariant way requires an accurate consideration of the dressed meson
propagators and photon emission vertices.

As stressed before, the approximations necessary to come to the
wave function at the origin prescription in case of the
positronium decay were justified, since only terms of higher orders
in $\alpha$ need to be neglected (such corrections can be taken into account systematically, see
\cite{pos2}).
In a strongly interacting system,
where the couplings are typically of order unity or larger,
these steps are not justified: they lead to uncontrolled results
and potentially large violations of gauge invariance.

\subsection{Case B: The zero--radius interaction limit}
\label{seczrl}

We now study the other limiting situation, case B
(Eq.~(\ref{schad})). In this limit we may assume the vertex
function to be point like
($\beta\to\infty$), which leads to a constant vertex function
$\varGamma(p,P)\equiv g_{S0}$ for, say, the decay of the $f_0$
into kaons.
Then all dressing effects can be absorbed in coupling constants
and masses and thus bare (in form!) vertices and propagators may
be used (but
for different reasons as compared to the previous subsection).

Then the matrix
element (\ref{vertex}) can be found from the set of diagrams
depicted in
Fig.~\ref{3d},
\be
\begin{array}{c}
\ds
W_a=\zeta g_{S0} e^2\int\frac{d^4k}{(2\pi)^4}\frac{\epsilon_1^*\cdot(2k-q_1)\;
\epsilon_2^*\cdot(2k-P-q_1)}{((k-q_1)^2-m^2)((k-P)^2-m^2)(k^2-m^2)},\\[5mm]
\ds W_b=W_a(1\leftrightarrow 2),\\[3mm]
\ds
W_c=-2 \zeta g_{S0} e^2(\epsilon_1^*\cdot\epsilon_2^*)\int\frac{d^4k}{(2\pi)^4}
\frac{1}{((k-P)^2-m^2)(k^2-m^2)},
\end{array}
\label{W1}
\ee
where $m$, as before, denotes the meson mass. 

The two-gamma decay of scalars can be viewed as a particular case of a
more general situation of the $S \to V \gamma$ decays, studied in
detail in Ref.~\cite{SVg}, with the vector particle $V$ also taken to be a
photon. The details of the calculations are well-known, and can be
found, {\em e.g.}, in Refs.~\cite{AI,NT,LP,BGP,CIK}.  Notice that,
although all integrals in Eq.~(\ref{W1}) are logarithmically
divergent, the sum $W_a+W_b+W_c$ is finite\footnote{An elegant way of
extracting the amplitude $M$, Eq.~(\ref{I}), by reading off a finite
coefficient at a specific combination of the four--momenta in $W_a$
was suggested in Ref.~\cite{CIK}.  The problem of convergence of the
integrals (\ref{W1}) was also studied in detail in
Refs.~\cite{AI,Markushin,moldec}}. Thus, adding these three yields for
the amplitude $M$ introduced in Eq.~(\ref{vertex}):
\begin{equation}
M(P^2)=-\zeta \frac{g_{S0}e^2}{2\pi^2m^2}I(b),
\label{I}
\end{equation}
with $I(b)$ being the loop integral function, $I(b)\equiv I(a=0,b)$ (see, for
example, Refs.~\cite{AI,CIK} for the definition of $I(a,b)$), where $b=m_S^2/m^2$,
\be
I(b)=\int_0^1 dz\int_0^{1-z}dy\,\frac{yz}{1-byz}.
\ee
The analytic expression for $I(b)$ takes the form:
\be I(b)=\left\{
\begin{array}{ll}
\ds-\frac{1}{2b}+\frac{2}{b^2}\left[\arcsin\frac{\sqrt{b}}{2}\right]^2,&b<4\\[2mm]
\ds-\frac{1}{2b}-\frac{1}{2b^2}\left[\ln\frac{\sqrt{b}+\sqrt{b-4}}{\sqrt{b}-\sqrt{b-4}}-i\pi
\right]^2,&b>4.
\end{array}
\right.
\label{I0}
\ee

Finally, using Eqs.~(\ref{observ}) and (\ref{I}) together, one arrives at the decay width
\be
\Gamma_{\gamma\gamma}^B=\zeta^2 \frac{g_{S0}^2\alpha^2 m_S^3}{16\pi^3m^4} \left|I\left(\frac{m_S^2}{m^2}\right)\right|^2,
\label{pw0}
\ee
with the only unknown parameter being the coupling constant $g_{S0}$.
For a loosely bound system with $P^2=(2m-\varepsilon)^2\approx 4m(m-\varepsilon)$,
$\varepsilon\ll m$, the condition (\ref{mnc}) gives the relation between the
coupling constant $g_{S0}$ and the molecule binding energy $\varepsilon$
\cite{moldec},
\be
\frac{g_{S0}^2}{4\pi}=32m\sqrt{m\varepsilon}.
\label{gSdef0}
\ee
Inserting Eq.~(\ref{gSdef0}) into Eq.~(\ref{pw0}) gives Eq.~(\ref{beauty}).

For $m_S=980$ MeV and $m=495$ MeV, which translates to
$\varepsilon=10$ MeV, one arrives at the prediction
\be
\Gamma_{\gamma\gamma}=0.22 \ \mbox{keV},
\label{pw2}
\ee
for the two--photon decay of the scalar $f_0(980)$, which we refer to as the point-like model prediction.
In the following we shall derive an estimate for the accuracy
of this result.

\section{Leading range corrections}
\label{seclrc}

In the previous subsection we investigated the limiting case of $\beta\to \infty$. In this chapter we derive the
leading corrections that emerge from finite values of $\beta$ --- we shall calculate the leading corrections
in $1/\beta$. This should provide a valuable insight into how accurate the formulae of the previous chapter should be expected to be.

For this we use a simple covariant model complying with the requirements of the previous chapter and thus providing a gauge invariant
description of the two--photon radiative decay of a non-point-like molecular state.

We start from an effective meson interaction Lagrangian which is responsible for the point-like
scalar formation and supply it with an extra momentum--dependent self--interaction: 
\be
L_{\rm int}=\frac12\lambda_1(\varphi^\dagger\varphi)^2+
\frac{\lambda_2}{2\beta^2}\left[\partial_\mu(\varphi^\dagger\varphi)\right]^2.
\label{Lint}
\ee
The form of the Lagrangian (\ref{Lint}) is chosen such that, after
inclusion of the e.m. field, it does not give rise to extra meson--photon
vertices. Indeed, since the Lagrangian (\ref{Lint}) is written completely in
terms of the real field $\varphi^\dagger\varphi$, the standard
substitution $\partial_\mu\to\partial_\mu-ieA_\mu$ does not touch it. As a
result, the set of diagrams contributing to the molecule decay to two photons
is not modified and is still exhausted with the three diagrams depicted in
Fig.~\ref{3d}. Additional terms that arise from possible charge exchanges
just lead to more complicated expressions but do not alter the conclusions.
This is discussed in detail in Appendix \ref{excur}.
The theory described by the Lagrangian (\ref{Lint}) can be renormalized to
the given order $1/\beta^2$. We present the necessary details in
Appendix \ref{renorm} and briefly summarize the results here.

The effective meson--meson interaction given rise by the Lagrangian (\ref{Lint}) is
\be
V(p-k)=\lambda_1+\frac{\lambda_2}{\beta^2}(p-k)^2.
\label{potV}
\ee
Note that, in addition to the two terms given in Eq.~(\ref{potV}) also a term that scales as
$\lambda_2(s/\beta^2)$ emerges from Eq.~(\ref{Lint}), where
$\sqrt{s}=E_{cm}$.
However, since we shall work at the fixed $s=m_S^2$, this term can be absorbed into $\lambda_1$.
The dressed meson propagator and the dressed photon emission vertex are
\be
S(p)=\frac{Z}{p^2-m^2},\quad v_\mu(p,q)=Z^{-1}(2p-q)_\mu+\tilde v_\mu (p,q),
\label{ren1}
\ee
where the renormalization factor $Z$ and the explicit expression for
$\tilde v_\mu (p,q)$ are given in Appendix \ref{renorm}. We have $\tilde v_\mu (p,q)q^\mu=0$, so that the Ward
identity (\ref{Wi}) is preserved. From now onwards we stick to the
renormalized, physical, value of the mass $m$. Besides that, $\tilde
v_\mu(p,q)$ does not contribute to the radiative $\gamma\gamma$ decay
under consideration since $(\tilde{v}\epsilon^*)_{q^2=0}=0$, with
$\epsilon^*_\mu$ being the photon polarization vector.

We turn now to the Bethe--Salpeter Eq.~(\ref{BS}) for a loosely bound system.
One can check that, to order $m^2/\beta^2$ and
$\sqrt{\varepsilon/m}$,
the Bethe--Salpeter Eq.~(\ref{BS}) is satisfied with the vertex function
(see Appendix \ref{renorm} for details):
\be
\varGamma(p,P)=Z^{-1}g_S\left(1+\xi\frac{p(p-P)}{\beta^2}\right),\quad\xi=\frac{\lambda_2}{\lambda_1},
\label{vGZ}
\ee
and the normalization condition (\ref{mnc}) gives (see Appendix \ref{norma} for the details):
\be
\frac{g_S^2}{4\pi}=32m\sqrt{m\varepsilon}\left(1+2\xi\frac{m^2}{\beta^2}\right),
\label{gSdef}
\ee which, as $\beta\to\infty$, reproduces the relation (\ref{gSdef0})
obtained in the limit of the zero--range interaction.

In the weak coupling limit that we focus on here, the bound state formation
should be controlled by non--relativistic momenta. As a consequence
$g_{\rm eff}$,
the effective coupling constant of the bound state to its constituents,
should have corrections at most of the order of 
$m\epsilon/\beta^2$~\cite{nonrelmol1,nonrelmol0},
for the scale $m^2$ does not appear in non--relativistic equations.
To recover this result we need to use
Eq.~(\ref{vGZ}) at the bound--state pole, $P=P_0$ with $P_0^2=m_s^2$, and for on--mass--shell mesons,
$p=p_0$ with $p_0^2=(p_0-P_0)^2=m^2$, to get
\begin{equation}
\frac{g_{\rm eff}^2}{4\pi}=\frac{Z^2}{4\pi}\varGamma^\dagger(p_0,P_0)\varGamma(p_0,P_0)
=32m\sqrt{m\varepsilon}\left(1+4\xi\frac{m\varepsilon}{\beta^2}\right) \ ,
\label{nonrelcoup}
\end{equation}
where the factor $Z^2$ was put according to the rules of the LSZ reduction formula.
The scaling of the corrections in Eq.~(\ref{nonrelcoup}) is in line with
the estimates of Refs.~\cite{nonrelmol1,nonrelmol0,moldec,nonrelmol2}.
Here we used that, for the given kinematics, $p(p-P)=(2m^2-m_s^2)/2$.

The general form of the matrix element is given in  Eq.~(\ref{vertex}).
Following the method proposed in Ref.~\cite{CIK} (see Appendix \ref{altern} for an alternative method) we notice that only
the diagrams (a) and (b) in Fig.~\ref{3d} give rise to the
structure $q_{1\nu}q_{2\mu}$ in the
transition matrix element (\ref{vertex}). Moreover, these two diagrams give
the same contribution to $W$, so it is sufficient to consider only one of
them,
\be
W_{\mu\nu}^{a}=e^2\zeta g_S\int\frac{d^4k}{(2\pi)^4}
\varGamma(k,P)S(k-P)v_{\nu}(k-q_1,q_2)S(k-q_1)v_{\mu}(k,q_1)S(k) \ ,
\ee
and to read off the coefficient at the structure $q_{1\nu}q_{2\mu}$ which
appears after the introduction of the Feynman parameters and shifting the
integration variable \cite{CIK}.
Notice that in this structure the $Z$--factors coming from
propagators, from e.m. vertices, and from the norm of the scalar vertex
cancel against each other, so that
\be
W_{\mu\nu}^{a}=\zeta g_Se^2 \int\frac{d^4k}{(2\pi)^4}
\frac{(2k-q_1)_\mu(2k-P-q_1)_\nu(1+\xi k(k-P)/\beta^2)}
{((k-q_1)^2-m^2)((k-P)^2-m^2)(k^2-m^2)}\ .
\ee

The corresponding loop integral is finite and the result reads:
\be
M(m_S^2)=M^{(0)}(m_S^2)+\xi\frac{m^2}{\beta^2}M^{(1)}
(m_S^2) \ ,
\label{Mfull}
\ee
where $M^{(0)}$ is given by the point-like result, Eq. (\ref{I}),
 whereas $M^{(1)}$ takes the form: \be
M^{(1)}(m_S^2)=-\zeta \frac{g_Se^2}{2\pi^2m^2}\tilde{I}(b),\quad
b=\frac{m_S^2}{m^2} \ ,
\label{M11}
\ee
with
\be
\tilde{I}(b)=\frac16-\frac12\int_0^1dz\int_0^{1-z}dy\left[(y+z-8yz)\ln(1-yzb)+\frac{1-y-z+2yz}{1-yzb}\right] \ .
\ee

Thus, up to order $1/\beta^2$, the integrals $I(b)$ and $\tilde{I}(b)$ enter the two--photon decay
amplitude in the combination $(I(b)+\tilde{I}(b))/I(b)$:
\be
M=M_{\rm point-like}\left(1+\xi\frac{m^2}{\beta^2}\frac{I(b)+\tilde{I}(b)}{I(b)}\right) \ ,
\label{mul1}
\ee
where the factor $1+\xi(m^2/\beta^2)$ comes from $g_S$ (see Eq.~(\ref{gSdef}))
and $I(b)+\xi(m^2/\beta^2)\tilde{I}(b)$ appears from the decay diagrams.
The integral $\tilde{I}(b)$ can be calculated analytically. The result
reads:
\be
\tilde{I}(b)=\left(1-\frac{b}{2}\right)I(b) \ .
\label{Ita}
\ee
So, for the total width, we arrive at an extremely simple formula:
\be
\Gamma_{\gamma\gamma}(\beta)=\Gamma_{\gamma\gamma}\left(1+\xi\frac{4m^2-m_S^2}{\beta^2}\right)
=\Gamma_{\gamma\gamma}\left(1+{\cal O}\left(\frac{m\varepsilon}{\beta^2}\right)\right) \ ,
\label{Gexact}
\ee
where $\Gamma_{\gamma\gamma}$ is given by Eqs.~(\ref{beauty}) and (\ref{pw2}),
and the corrections of order of $m^2/\beta^2$ cancel against each other.
Contrary to Eq.~(\ref{nonrelcoup}),
here the cancellation is unexpected and non--trivial: since
the photons carry away an energy of the order of the mass $m$,
their momenta are the same and therefore at least one of
the particles in the meson loop has a typical momentum of
the order of its mass. Consequently there is no justification
for the use of non--relativistic kinematics in the evaluation
of the two--photon decay of scalar mesons. 

Evaluation of the actual coefficient in front of the structure
$(m\varepsilon)/\beta^2$ would require making assumptions concerning the
details of the molecule formation which are model--dependent, though, given a
particular model of this type, it is straightforward to apply the technique of
the present work to establish this coefficient which is expected to be of order unity (see also Appendix \ref{norma}).

Equation (\ref{Gexact}) is the central result of this work, for it
shows that the predictions derived for the limit a point-like
interaction should be quite accurate. If one assumes the coefficient
$\xi$ also to take its natural value of order unity
($\lambda_1\sim\lambda_2$), we find that the leading range corrections to Eq.~(\ref{pw2}) should be
of the order of $m\varepsilon/\beta^2$, which translates into a few percent
 in the decay amplitude. Therefore the accuracy of Eq.~(\ref{beauty}) should
be given by the sub-leading range corrections that are expected to be of
the order of $(m/\beta)^4$, which is about 15\% for the case of the $f_0$.

\begin{figure}[t]
\begin{center}
\epsfig{file=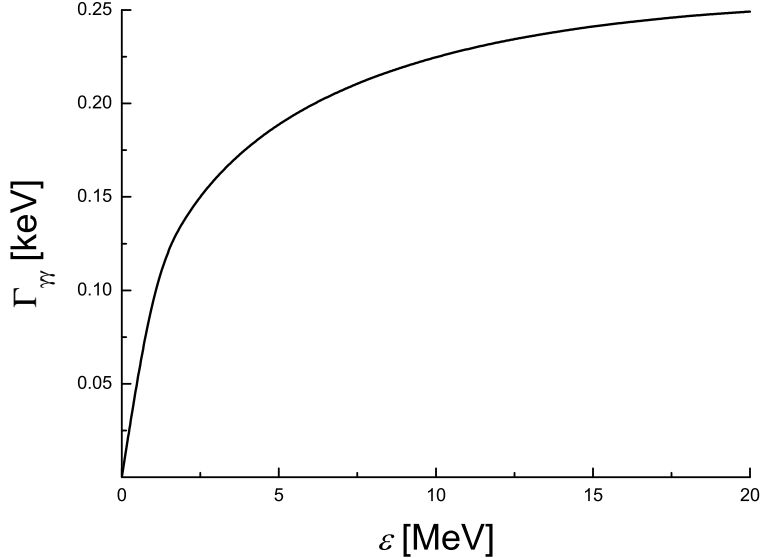,width=10cm}
\end{center}
\caption{Dependence of the width $\Gamma_{\gamma\gamma}$, defined in Eq.~(\ref{beauty}), on
the value of the binding energy.}
\label{epsdep}
\end{figure}

Another source of uncertainty for any prediction based on
Eq.~(\ref{beauty}) is our ignorance on the true binding energy. To
investigate this point, in Fig.~\ref{epsdep}, the dependence of
Eq.~(\ref{beauty}) on $\varepsilon$ is shown, again using for illustration the
parameters relevant for the $f_0$, namely $m=m_K$. As one can see, the
dependence on $\varepsilon$ is quite moderate, once the binding energy
exceeds $5$ MeV. Therefore, even if $\varepsilon$ is varied between $5$
and $20$ MeV around $10$ MeV --- the typical value used above --- the
predicted two--photon width changes by less than 0.05 keV.

However, one should be aware of the following important {\it disclaimers}:
\begin{itemize}
\item most of the hadrons --- including the $f_0$ --- are unstable. Thus the
concept of vertex function and binding energy is not well defined for those, and one should
employ a multi-channel Bethe--Salpeter formalism.
The quantity that
should replace the bound--state vertex $\varGamma(p,P)$ in all the formulae given above
is the multi-channel $t$-matrix. The proof of gauge invariance proceeds along the lines
similar to those given in Section~\ref{bsa}. In the molecular case, for the energies around
the $K \bar K$ threshold (and far away from the inelastic thresholds) the amplitude
in the $K \bar K$ channel can be written in the scattering length approximation with the
complex $K \bar K$ scattering length:
\be
a_{K \bar K}=\frac{1}{\kappa_1+i\kappa_2},\quad\kappa_2>0.
\ee
In the limit $\kappa_2 \to 0$, the coupling to inelastic channels is switched off,
and, for $\kappa_1>0$, there is a bound state in the $K \bar K$ channel with
$\varepsilon=\kappa_1^2/m$. As shown in
\cite{Flatte}, the data on, say, $\pi\pi$ scattering near the $K \bar K$ threshold
can be described in the scattering length approximation with $\kappa_2$ around $50\div 100$
MeV, and the ratio $\kappa_1/\kappa_2$ of order unity. Thus, the hierarchy of scales in the case of unstable
scalar is similar to the one considered above. The two--photon decay of an unstable scalar 
meson in the limit of  point-like interactions was evaluated, for example, in Ref.~\cite{ach2}.
More systematic studies of the problem of unstable particles will be subject of a future
work.
\item Another issue is the possible presence of additional short--ranged operators, for example, of the type of vector meson
exchanges studied in Ref.~\cite{OO}. Estimates for these and their proper inclusion in the renormalization
program also go beyond the scope of the present paper and will also be subject of a future work.
\end{itemize}

\section{Summary}

\begin{enumerate}
\item The $\Psi(0)$ formula for slow particles annihilation does not work for the two--gamma decays of
hadronic molecules. Not only are the results numerically uncontrolled,
which is reflected in a wide spread of predictions for the decay $f_0\to \gamma\gamma$ width found in the literature,
but there is also a potentially
large violation of gauge invariance necessarily present in the derivation
of the formula.
\item 
Simple recipes to restore gauge invariance in the presence of non--trivial
vertex functions through new contact diagrams with the derivatives of
this vertex, successfully used for decays like $\phi\to \gamma f_0$~\cite{moldec}, fail, since the photons are
not soft. 
As a result, gauge invariance, preserved in the point-like limit, appears broken already to order $1/\beta^2$.
We showed that the inclusion of the scalar vertex structure in a gauge
invariant way requires an accurate consideration of the dressed meson
propagators and photon emission vertices.
\item For phenomenologically adequate values of $\varepsilon=10$ MeV and $\beta\sim 0.8$ GeV
for the scalar meson $f_0(980)$ our prediction for the two--photon width is
\be
\Gamma_{\gamma\gamma}^{f_0} = (0.22\pm 0.07)\ \mbox{keV}.
\label{ourpred}
\ee
Our result compares nicely with the experimental values for the $\gamma\gamma$ width of
the light scalar $f_0(980)$~\cite{PDG}
\begin{equation}
\Gamma_{\gamma \gamma}(f_0(980))=0.31^{+0.08}_{-0.11}~{\rm keV},
\label{PDG}
\end{equation}
and~\cite{Pen}
\begin{equation}
\Gamma_{\gamma \gamma}(f_0(980))=0.28^{+0.09}_{-0.13}~{\rm keV}.
\label{Pen}
\end{equation}
The new experimental value \cite{newdata}
\be
\Gamma_{\gamma \gamma}(f_0(980))=0.205^{+0.095}_{-0.083}({\rm stat})^{+0.147}_{-0.117}({\rm syst})~{\rm keV}
\label{newbelle}
\ee
gives an even better agreement. This clearly supports the molecular assignment for the $f_0(980)$.
\end{enumerate}
It has to be stressed that the uncertainty of our theoretical
prediction (\ref{ourpred}) so far only includes our estimate of the
possible influence of the structure of the vertex function for the
scalar meson (about 15\% for the amplitude).  Neither was the possible
influence of the finite width included nor possible additional terms
from shorter ranged transitions. Both will be subject of future
investigations.

It should be emphasized that the main goal of our study was to quantify
the effect of range corrections to the two--photon decay of hadronic
molecules in a model independent way. Those we identified as
parametrically suppressed compared to what is expected naively. This
finding should not be changed by the inclusion of inelastic channels
(like $\pi\pi$ in case of the $f_0$). From this point of view our work
is an additional justification for the use of, for example, the chiral unitary
approach, for the calculation of the two--photon decay of the light
scalar mesons~\cite{OO}. Here scalar mesons appear as hadronic molecules based
on point--like interactions. On the other hand, in Ref.~\cite{OO} the 
$\pi\pi$ channel is included in a coupled channel framework. It is
also reassuring that the width calculated in this reference is consistent
with our result, Eq.~(\ref{ourpred}).

\vspace{0.2cm}

{\bf Acknowledgments}

This research was supported by the Federal Agency for Atomic Energy of Russian Federation,
by the grants RFFI-05-02-04012-NNIOa, DFG-436 RUS 113/820/0-1(R), NSh-843.2006.2, and NSh-5603.2006.2,
by the Federal Programme of the Russian Ministry of Industry, Science, and
Technology No. 40.052.1.1.1112, and by the Russian Governmental Agreement
N 02.434.11.7091. A.N. is also supported through the project PTDC/FIS/70843/2006-Fisica.

\appendix

\section{Renormalization of the model and the Bethe--Salpeter equation}
\label{renorm}

To renormalize the theory (\ref{Lint}) to the given order $1/\beta^2$ we start from the interaction Lagrangian
\be
L_{int}=\frac12 \lambda_1(\varphi^\dagger\varphi)^2+
\frac{\lambda_2}{2\beta^2}\left[\partial_\mu(\varphi^\dagger\varphi)\right]^2.
\label{Lint2}
\ee

We consider the meson mass operator then,
\be
\Sigma(p)=-i\int\frac{d^4k}{(2\pi)^4}S(k)V(p-k),
\ee
and use the dimensional regularization scheme to make it finite. It is easy to see that $\Sigma(p)$ can be written in the form
\be
\Sigma(p)=(1-Z^{-1})p^2+\delta m^2,
\ee
where
\be
Z=1+\frac{\lambda_2m^2}{(4\pi)^2\beta^2}(\Lambda+1),\quad \Lambda=\frac{2}{\epsilon}-\gamma_E+\ln\frac{4\pi\mu^2}{m^2},
\label{LL}
\ee
with $D=4-\epsilon$ being the number of dimensions, $\mu$ and $\gamma_E\approx 0.577$ being an auxiliary mass parameter and the Euler constant, respectively.
The physical meson mass is simply $m^2=Z(m_0^2+\delta m^2)$, and the meson
propagator takes the form given in Eq.~(\ref{ren1}).
It is also straightforward to evaluate, to the same order $1/\beta^2$ and in the same regularization scheme, the photon emission vertex,
\be
v_\mu(p,q)=(2p-q)_\mu-i\int\frac{d^4k}{(2\pi)^4}V(p-k)S(k)S(k-q)(2k-q)_\mu,
\ee
to arrive at
\be
v_\mu(p,q)=Z^{-1}(2p-q)_\mu + \tilde v_\mu (p,q) \ ,
\ee
where
\begin{equation}
\tilde v_\mu (p,q)=\frac{\lambda_2}{12\pi\beta^2}\left[\frac{\Lambda}{3}+\frac16-\frac{4m^2-q^2}{q^2}
+\left(\frac{4m^2-q^2}{q^2}\right)^{3/2}\arctan\sqrt{\frac{q^2}{4m^2-q^2}}\right][q^2p_\mu-(pq)q_\mu],
\end{equation}
with the renormalization factor $Z$ given in Eq.~(\ref{LL}). This agrees with Eq.~(\ref{ren1}).

Now, before we come to the Bethe--Salpeter equation, we introduce two auxiliary integrals,
\begin{eqnarray}
I_0(P)=-i\int\frac{d^4k}{(2\pi)^4}\frac{1}{(k^2-m^2)((k-P)^2-m^2)}\nonumber,\\[-3mm]
\\
I_2(P)=-i\int\frac{d^4k}{(2\pi)^4}\frac{k^2}{(k^2-m^2)((k-P)^2-m^2)},\nonumber
\end{eqnarray}
which are divergent and, in the dimensional regularization scheme, take the form:
\be
I_0(P)=\frac{1}{(4\pi)^2}\Lambda +I_{0R}(P),\quad
I_2(P)=\frac{2m^2}{(4\pi)^2}\Lambda+I_{2R}(P),
\ee
where $I_{0R}$ and
$I_{2R}$ are finite.

The Bethe--Salpeter equation is
\be
\varGamma(p,P)=-i\int\frac{d^4k}{(2\pi)^4} \varGamma(k,P)S(k)S(k-P)\left(\lambda_1+\frac{\lambda_2}{\beta^2}(p-k)^2\right),
\ee
where dressed kaon propagators should be used.
In the leading order in $1/\beta^2$ the scalar vertex can be found in the form:
\be
\varGamma(p,P)=g_1+\frac{g_2}{\beta^2}p(p-P),
\ee
with the coefficients $g_1$ and $g_2$ satisfying the equations:
\be
g_1=\lambda_1g_1I_0(P)+2\frac{\lambda_1\lambda_2g_1}{\beta^2}I_0(P)\frac{m^2}{(4\pi)^2}(\Lambda+1)+
\frac{\lambda_1g_2}{\beta^2}I_2(P)
-\frac{\lambda_1g_2P^2}{2\beta^2}I_0(P)+\frac{\lambda_2g_1}{\beta^2}I_2(P),
\ee
\be
g_2=\lambda_2g_1I_0(P)\label{g2}.
\ee

These yield the equation which defines the mass $M$ of the bound state as ($P^2=M^2$)
\be
1-\lambda_1I_0(P)-\frac{\lambda_2}{\beta^2}\left[\lambda_1I_0(P)\frac{2m^2}{(4\pi)^2}(\Lambda+1)+
 \lambda_1 I_0(P)I_2(P)-
\frac{1}{2}P^2\lambda_1I_0^2(P)+I_2(P)\right]=0.
\label{boundstate}
\ee
We treat this equation perturbatively in $1/\beta^2$. Then, in the zeroth order, one has
\be
\frac{1}{\lambda_1}=I_0(M_0),
\label{zero}
\ee
where $M_0$ is zero--order mass of the bound state. The divergent part of the integral $I_0(M_0)$ is absorbed into the coupling constant $\lambda_1$. Then, in the next--to--leading order,
\be
\frac{1}{\lambda_1}-I_0(P)-\frac{2\xi}{\beta^2}\left[\frac{m^2}{(4\pi)^2}(\Lambda+1)+I_2(P)-\frac{1}{4}P^2I_0(P)\right]=0,\quad\xi=\lambda_2/\lambda_1,
\ee
with the problem of renormalization solved similarly to Eq.~(\ref{zero}). As $g_2$ enters the vertex together with $1/\beta^2$, Eqs.~(\ref{g2}) and (\ref{zero}) together yield
\be
g_2=\xi g_1,
\ee
and thus we arrive at the vertex function in the form
\be
\varGamma(p,P)=g_1\left(1+\xi\frac{p(p-P)}{\beta^2}\right),
\label{g1g2}
\ee
which requires normalization. This is discussed in detail in Appendix \ref{norma}.

\section{Normalization of the vertex function}
\label{norma}

Normalization of the vertex function is given by Eq.~(\ref{mnc}). In the zeroth order in the $1/\beta^2$ expansion this gives
\be
g_1^2\frac{\partial I_0(P)}{\partial P^2}=1,
\label{normzero}
\ee which, for a loosely bound state with $\varepsilon\ll m$,
reproduces the relation (\ref{gSdef0}) with $g_{S0}=g_1$. Let us go
beyond the zeroth order now and include corrections $\propto
1/\beta^2$. The form of $\varGamma(p,P)$ given in
Eq.~(\ref{g1g2}) obviously represents the first two terms in the
successive expansion of the exact vertex function in the inverse
powers of $\beta^2$. Such an expansion performed prior to taking
integrals with $\varGamma(p,P)$ involved may explode if the rest of the
integrand does not converge fast enough to suppress the contribution
of the higher and higher powers of the loop momentum which appear in
the $1/\beta^2$ expansion of $\varGamma(p,P)$. This is obviously not the
case for the loop integral (\ref{mnc}) and thus we face the problem of
convergence of the normalization integral already to order
$1/\beta^2$. Notice, however, that the solution of this problem is
well--known --- the normalization integral evaluated with the vertex
function in the full form converges, and the expansion in $1/\beta^2$
is to be performed afterwards. Building the full form
of the vertex function would require resorting to a particular model
of the molecule formation which we would like to avoid in our general
consideration. Fortunately, for loosely bound states, the
model--dependent contributions to the vertex function appear in higher
orders in the $\varepsilon/m$ expansion. Indeed, if we substitute the
vertex function (\ref{g1g2}) to the normalization condition
(\ref{mnc}) and perform integration in $d^4k$ retaining only the terms
of order $\sqrt{\varepsilon/m}$ and neglecting all higher
contributions, then the result can be expressed entirely through the
integrals $I_0(P)$ and $I_2(P)$ defined in Appendix \ref{renorm}, computed to the
same order, 
\be
I_0(P)=\frac{1}{(4\pi)^2}\Lambda+\frac{1}{8\pi^2}-\frac{1}{16\pi}\sqrt{\frac{\varepsilon}{m}}+\ldots,\quad
I_2(P)=\frac{2m^2}{(4\pi)^2}\Lambda+\frac{3m^2}{16\pi^2}-\frac{m^2}{16\pi}\sqrt{\frac{\varepsilon}{m}}+\ldots.
\ee 
All divergent contributions disappear, as discussed in Appendix \ref{renorm}, and the result (\ref{gSdef})
is readily reproduced, where \be g_S=Zg_1.  \ee Then, finally, the
scalar vertex function takes the form of Eq.~(\ref{vGZ}).

As a cross--check of the results (\ref{vGZ}) and (\ref{gSdef}) we assume a particular form of the full scalar vertex compatible with the large--$\beta$ expansion (\ref{vGZ}). As such we choose, for the sake of transparency, the form:
\be
\varGamma(p,P)=Z^{-1}g_S\frac{\beta^2}{\beta^2-\xi p(p-P)}.
\label{G2}
\ee
Then, introducing Feynman parameters and integrating out the four--momentum, one can rewrite the normalization condition (\ref{mnc})
in the form:
\be
\frac{g_S^2}{8\pi^2}\int_0^1 \frac{zdz}{Q^2}\int_0^{(1-z)\beta^2/(\xi Q^2)}
dx\frac{x(1-z-\xi xQ^2/(2\beta^2))}{[x^2\xi^2 Q^2P^2/\beta^4+x(1+\xi(zP^2-m^2)/\beta^2)+1]^3}=1,
\label{norm2}
\ee
where $Q^2=m^2-z(1-z)P^2$. The integral in $x$ can be easily evaluated and yields, to order $1/\beta^2$,
\be
\int_0^{(1-z)\beta^2/(\xi Q^2)} dx\frac{x(1-z-\xi x Q^2/(2\beta^2))}{[x^2\xi^2
Q^2P^2/\beta^4+x(1+\xi(zP^2-m^2)/\beta^2)+1]^3}
\approx\frac12(1-z)\left(1-2\xi\frac{zP^2-m^2}{\beta^2}\right),
\ee
so that the relation (\ref{norm2}) reduces to
\be
\left(1-\xi\frac{P^2-2m^2}{\beta^2}\right)\frac{g_S^2}{16\pi^2}\int_0^1 dz\frac{z(1-z)}{Q^2}=1,
\label{norm3}
\ee
where the symmetry of the function $z(1-z)/Q^2$ with respect to the variable change $z\to 1-z$ was used.
The remaining integral in $z$ is specific for the point-like limit and it was evaluated before in order to derive the relation (\ref{gSdef}). Therefore, one can rewrite (\ref{norm3}) in the form
\be
\frac{g_S^2}{4\pi}=\left(1+\xi\frac{P^2-2m^2}{\beta^2}\right)\left(\frac{g_S^2}{4\pi}\right)_{\rm point-like}
=\left(1+2\xi\frac{m^2}{\beta^2}+{\cal O}\left(\frac{m\varepsilon}{\beta^2}\right)\right)\left(\frac{g_S^2}{4\pi}\right)_{\rm point-like}.
\ee
Thus the relation (\ref{gSdef}) is re-derived plus the first correction of the form $(m\varepsilon)/\beta^2$ is 
established for the vertex function (\ref{G2}). As it was anticipated before, the term of order $m^2/\beta^2$ 
is model--independent 
and coincides with the one obtained in the simple approach described in the beginning of this appendix.

The result of this appendix can be understood in the language of effective field theories. Indeed, 
all divergencies are to be absorbed into appropriate
counter terms, however, there are no counter terms allowed that are
non--analytic in $\varepsilon$. Consequently all terms that scale as
$\sqrt{\varepsilon}$ are fixed model independently.

\section{Failure of a simple recipe of gauge invariance restoration for a non--trivial vertex function $\varGamma(k,P)$}
\label{fail}

In this appendix we demonstrate that gauge invariance appears broken to order $m^2/\beta^2$ for the naive attempts to consider a non--trivial vertex function $\varGamma(k,P)$
without a self--consistent dressing of the photon emission vertices and the meson propagators. Thus we start from Eq.~(\ref{WG}) with a non--trivial vertex function
$\varGamma(k,P)$ but with the bare photon emission vertices and meson propagators. This yields:
\be
W_{\mu\nu}q_1^{\mu}=e^2\int\frac{d^4k}{(2\pi)^4}[\varGamma(k+q_1,P)-\varGamma(k,P)]S(k)S(k-q_2)(2k-q_2)_\nu.
\ee
For $\varGamma(k,P)\neq$const the difference in the square brackets does not vanish, and 
as a result also $W_{\mu\nu}q_1^{\mu}$ remains finite. 
A naive counting of powers of $\beta$ shows that this difference scales as $1/\beta^2$ (see, for example, Eq.~(\ref{vGZ}) for the form of
the vertex $\varGamma(k,P)$) and thus a 
simple trick with 
adding contact diagrams with the photon emission from the scalar vertex described by $\partial\varGamma/\partial k^\mu$ 
could solve the problem to the given order $1/\beta^2$, so
that gauge invariance breaking happens only in the order $1/\beta^4$ which we neglect throughout the paper. 
In Ref.~\cite{moldec} this approach was successfully used for the decay $\phi\to\gamma f_0$
--- the interested reader can find the details and the discussion of the method, for example, 
in the aforementioned paper. 
In contrast to the $\phi$--decay, here we have two identical photons in the final
state.
The requirement of symmetry of the amplitude with respect to the 
interchange of these leads to two single contact diagrams and one double contact diagram, the latter containing 
$\partial^2\varGamma/\partial k^\mu\partial k^\nu$ in the scalar vertex. The  contributions to be added to $W_{\mu\nu}q_1^{\mu}$
take the form:
\begin{eqnarray}
\ds \delta W_{\mu\nu}^{(1)}q_1^{\mu}&=&-q_1^\mu e^2\int\frac{d^4k}{(2\pi)^4}\frac{\partial\varGamma(k,P)}{\partial k^\mu}S(k)S(k-q_2)(2k-q_2)_\nu,\nonumber\\[1mm]
\ds \delta W_{\mu\nu}^{(2)}q_1^{\mu}&=&-q_1^\mu e^2\int\frac{d^4k}{(2\pi)^4}\frac{\partial\varGamma(k,P)}{\partial k^\nu}S(k)S(k-q_1)(2k-q_1)_\mu,\label{Wqq0}\\[1mm]
\ds \delta W_{\mu\nu}^{(3)}q_1^{\mu}&=&q_1^\mu e^2\int\frac{d^4k}{(2\pi)^4}\frac{\partial^2\varGamma(k,P)}{\partial k^\mu\partial k^\nu}S(k),\label{naiveterms}
\end{eqnarray}
so that the resulting expression reads
\begin{eqnarray}
\ds W_{\mu\nu}q_1^{\mu}=e^2\int\frac{d^4k}{(2\pi)^4}\left[\varGamma(k+q_1,P)-\varGamma(k,P)-q_1^\mu
\frac{\partial\varGamma(k,P)}{\partial k^\mu}\right]\hspace*{4cm}\nonumber\\
\label{Wqq}\\
\hspace*{8cm}\ds\times\left[(2k-q_2)_\nu S(k-q_2)S(k)-2k_\nu S^2(k)\right]\nonumber,
\end{eqnarray}
where the integral coming from the double contact vertex was integrated by parts. 
In order to proceed we use the form of Eq.~(\ref{G2}) for the vertex
function --- since we are only after the scaling behaviour of the remaining
violation of gauge invariance, here we are free to work within a particular
model. This choice ensures convergence of the loop integral and complies with the large--$\beta$ expansion (\ref{vGZ}). Then we get
\be
\varGamma(k+q_1,P)-\varGamma(k,P)-q_1^\mu\frac{\partial\varGamma(k,P)}{\partial k^\mu}=
Z^{-1}g_S\frac{\beta^2\xi^2[2(kq_1)-(q_1q_2)]^2}{[\beta^2-\xi(k+q_1)(k-q_2)]^2[\beta^2-\xi k(k-q_1-q_2)]},
\label{GGG}
\ee
which scales as $1/\beta^4$ as $\beta\to\infty$. However, in contrast to this the corresponding integral of  Eq.~(\ref{Wqq}) for
$W_{\mu\nu}q_1^{\mu}$ scales as $1/\beta^2$.  To see this observe that convergence
to the integral is provided by the denominator with the consequence that $k$ takes 
values of the order of $\beta$ (for large $\beta$). Therefore, the relevant estimate for Eq.~(\ref{GGG}) is 
$\beta^2(kq_1)^2/\beta^6\sim \omega^2/\beta^2$, where $\omega$ is the typical
energy of a photon. We checked by an explicit calculation that this behaviour is valid for the integral (\ref{Wqq}) as well.

Therefore, even introducing the correction terms of Eqs. (\ref{naiveterms}),
does not change the order where a violation of gauge invariance appears --- it 
appears at order $\omega^2/\beta^2$.
In the given kinematics photons are not soft, $\omega\approx m$. Thus the simple prescription described in this appendix
to cure the violation of gauge invariance does not improve the situation,
since gauge invariance is still broken by the terms of the 
order of $m^2/\beta^2$. 
  
\section{Alternative derivation of Eq. (\ref{M11})}
\label{altern}

As a cross--check of gauge invariance, let us extract the amplitude
(\ref{M11}) from the coefficient at the structure $g_{\mu\nu}$
in Eq.~(\ref{vertex}). This is less trivial as the seagull diagram (Fig.~\ref{3d}(c))
contributes. This seagull has $Z^{-1}$ factor coming from the scalar vertex
and $Z^2$ factor due to the two meson propagators. Because of this mismatch
of $Z$-factors, in addition to the contribution giving the result (\ref{M11}),
a divergent piece arises in $W_{\mu\nu}$, which comes from the leading
term, of order $(1/\beta^2)^0$, in the scalar vertex $\varGamma(p,P)$:
\be
-2\zeta ig_{\mu\nu}g_Se^2 I_0(Z-1)=-2\zeta ig_{\mu\nu}g_Se^2
I_0\frac{\lambda_2m^2}{\beta^2(4\pi)^2}(\Lambda+1),
\label{seagull}
\ee
where the expressions for the $Z$ and $\Lambda$ in the dimensional
regularization scheme are given in Eq.~(\ref{LL}). 
To order $1/\beta^2$, one can make use of Eq.~(\ref{zero}) and replace $\lambda_2I_0$ in (\ref{seagull}) by
\be
\lambda_2I_0=\xi\lambda_1I_0=\xi.
\ee

In the meantime,
another divergent contribution comes from the $1/\beta^2$ term in $\varGamma(k,P)$, which reads:
\be
2\zeta ig_{\mu\nu} g_Se^2
\frac{\xi}{\beta^2}\frac{m^2}{(4\pi)^2}(\Lambda+1),
\label{first}
\ee
so that the two undesired divergent contributions to $W_{\mu\nu}^a$ cancel against each other and the
gauge--invariant formula (\ref{vertex}) is re-derived,
with $M(P^2)$ given
by Eq.~(\ref{Mfull}).

\section{Account for exchange currents}
\label{excur}

The interaction Lagrangian (\ref{Lint}) does not give rise to extra
kaon--photon vertices, in addition to those following from the kinetic
part of the kaons Lagrangian. In this appendix we consider the
possibility for these vertices to appear due to meson exchange
currents.

Let us introduce field doublet $\vp^{\alpha}$, and have the interaction
Lagrangian in the form
\be
L_{int}=\frac12 \lambda_1(\vp^+\vt\vp)^2 + \frac{\lambda_2}{2\beta^2}[\partial_{\mu}(\vp^+\vt\vp)]^2.
\ee
In momentum space, it gives rise to the four--point vertex of the form
\be
\left(\lambda_1+\frac{\lambda_2}{\beta^2}(p-k)^2\right)\vt^{\alpha}_{\beta}\vt^{\lambda}_{\rho},
\ee
where $p$ and $k$ are the momenta of the incoming and outgoing kaons, and $\alpha$, $\beta$ and $\rho$, $\lambda$
are the isospin indices of incoming and outgoing mesons, respectively.
As before, the term linear in $s$ was absorbed into $\lambda_1$.

Electromagnetic interaction is then found from the minimal substitution,
\be
p_{\mu}\delta^{\alpha}_{\beta} \to p_{\mu}\delta^{\alpha}_{\beta}-eA_{\mu}Q^{\alpha}_{\beta},
\ee
where $Q=(1+\tau_3)/2$ is the charge operator. Thus two new kaon--photon vertices are generated in the order $1/\beta^2$:
the contact single--photon vertex ($q_\mu$ is the photon momentum),
\be
i\frac{\lambda_2}{\beta^2}
\epsilon_{3kn}\tau^{\alpha}_{k\beta}\tau^{\lambda}_{n\rho}(2k-2p+q)_{\mu},
\label{single}
\ee
and the double--photon vertex,
\be
2g_{\mu\nu}\frac{\lambda_2}{\beta^2}(\vt^{\alpha}_{\beta}\vt^{\lambda}_{\rho}-
\tau^{\alpha}_{3\beta}\tau^{\lambda}_{3\rho}).
\label{double}
\ee

The dressed propagator is now
\be
S^{\alpha}_{\beta}=\delta^{\alpha}_{\beta}S(p),~~S(p)=\frac{Z}{p^2-m^2},
\label{S}
\ee
with
\be
Z=1+\frac{3\lambda_2m^2}{(4\pi)^2\beta^2}(\Lambda+1).
\label{Z}
\ee

The dressed photon emission vertex,
\be
\begin{array}{c}
\ds v^{\lambda}_{\mu,\beta}(p,q)=Q^{\lambda}_{\beta}(2p-q)_{\mu}-i\frac{\lambda_2}{\beta^2}\vt^{\lambda}_{\rho}
Q^{\rho}_{\alpha}\vt^{\alpha}_{\beta}\int\frac{d^4k}{(2\pi)^4}S(k)S(k-q)(p-k)^2(2k-q)_{\mu}\\
\ds +\frac{\lambda_2}{\beta^2}\epsilon_{3kn}\tau^{\alpha}_{\beta}\tau^{\lambda}
_{\alpha}\int
\frac{d^4k}{(2\pi)^4}S(k)(2k-2p+q)_{\mu},
\end{array}
\label{v}
\ee
satisfies the Ward identity
\be
q^{\mu}v^{\lambda}_{\mu,\beta}=Q^{\lambda}_{\beta}[S^{-1}(p)-S^{-1}(p-q)],
\label{ward}
\ee
with $S(p)$ given by Eqs.~(\ref{S}) and (\ref{Z}). After simple algebraic transformations the photon emission vertex (\ref{v}) can be represented as
\be
v^{\lambda}_{\mu,\beta}(p,q)=Q^{\lambda}_{\beta}Z^{-1}(2p-q)_{\mu}+\ldots,
\ee
where the ellipsis denotes the terms irrelevant to the $\gamma\gamma$ decay
involving real photons.

The Bethe--Salpeter equation is now
\be
\varGamma^{\alpha}_{\rho}(p,P)=-i\int\frac{d^4k}{(2\pi)^4}\vt^{\alpha}_{\beta}\varGamma^{\beta}_{\lambda}(k,P)
\vt^{\lambda}_{\rho}S(k)S(k-P)\left(\lambda_1+\frac{\lambda_2}{\beta^2}(p-k)^2\right).
\ee

The momentum dependence of the vertex and the normalization condition are the same as in Eqs.~(\ref{vGZ}) and (\ref{gSdef}), 
respectively, while the matrix structure of the vertex is either
\be
\varGamma^{\alpha}_{\rho}(p,P)\propto\frac{1}{\sqrt{2}}\delta^{\alpha}_{\rho},
\label{singlet}
\ee
in the isosinglet case, or
\be
\varGamma^{\alpha}_{\rho}(p,P)\propto\frac{1}{\sqrt{2}}\tau^{\alpha}_{k\rho},
\label{triplet}
\ee
in the isotriplet one. Consequently, in the former case, Eq.~(\ref{zero}) is replaced by
\be
\frac{1}{\lambda_1}=3I_0(P),
\label{Isinglet}
\ee
whereas, in the latter case, it becomes
\be
\frac{1}{\lambda_1}=-I_0(P).
\label{Itriplet}
\ee
Thus one may have either a isosinglet or a isotriplet bound state, depending on the sign of $\lambda_1$.

\begin{figure}[t]
\begin{center}
\begin{tabular}{ccc}
\raisebox{0mm}{\epsfig{file=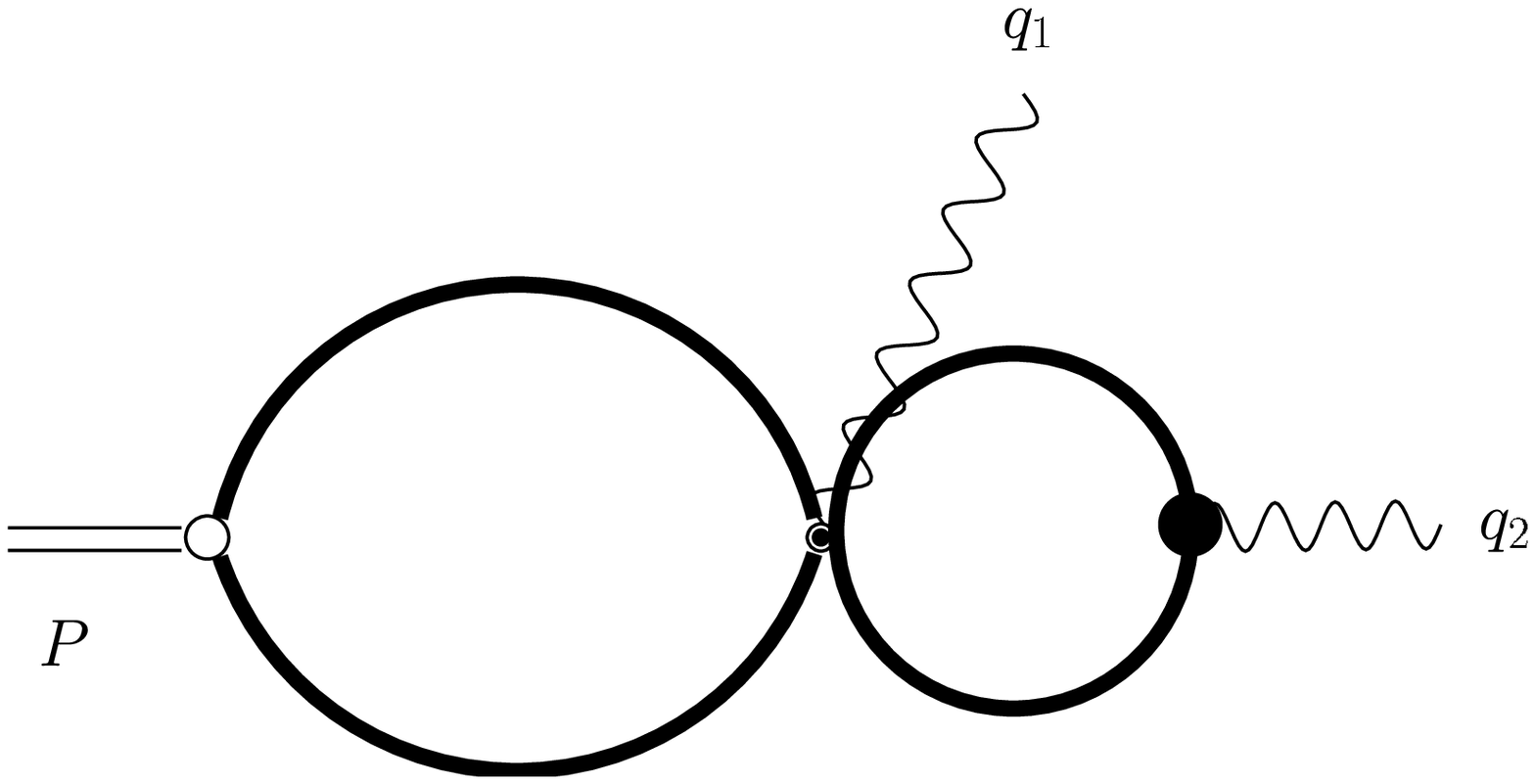,width=7cm}}&\hspace*{1cm}&\raisebox{-12mm}{\epsfig{file=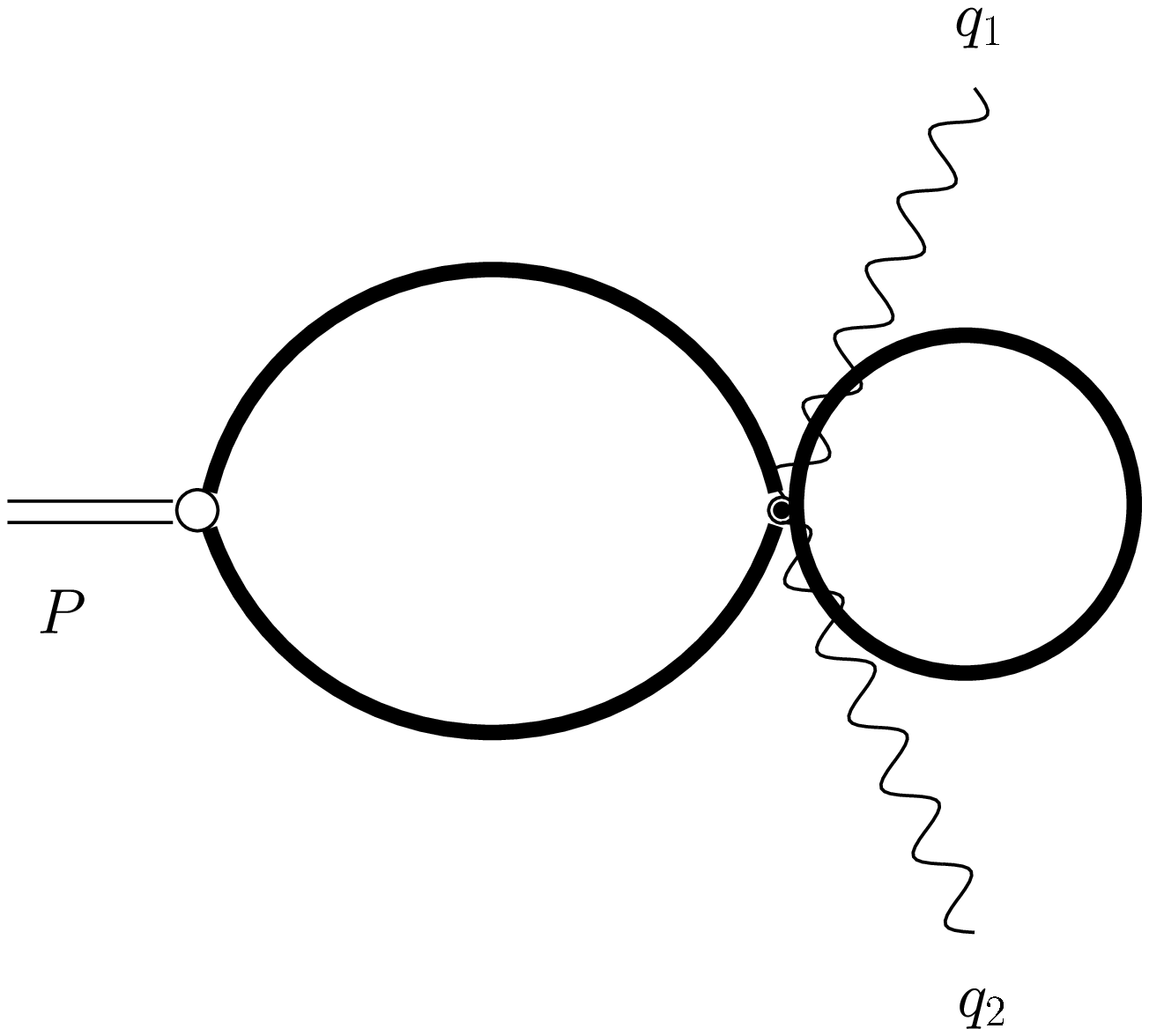,width=5cm}}\\
(d)&&(e)\\
\end{tabular}
\end{center}
\caption{Extra diagrams contributing to the two--photon scalar decay amplitude due to exchange currents.}\label{2exd}
\end{figure}

Let us turn to the calculation of the two--gamma decay amplitude. The contributions to the decay amplitude
proportional to $q_{1\nu}q_{2\mu}$ are left intact by inclusion of the exchange currents and they are given by the
graphs (a) and (b) of the Fig.~\ref{3d}. In the meantime, extra terms appear which contribute to the coefficient at the 
structure $g_{\mu\nu}$. Because of the mismatch of $Z$--factors, the divergent piece coming from the leading order term of
$\varGamma^{\alpha}_{\rho}(p,P)$ appears, similarly to Eq.~(\ref{seagull}), to be
\be
-6ig_{\mu\nu}\frac{g_Se^2}{\sqrt{2}}I_0\lambda_1\frac{\xi}{\beta^2}\frac{m^2}
{(4\pi)^2}(\Lambda+1),
\ee
while the contribution from the $1/\beta^2$ term in $\varGamma^{\alpha}_{\rho}(p,P)$
remains the same as in the neutral exchange case --- see Eq.~(\ref{first}). Thus, using the relations (\ref{Isinglet}) and (\ref{Itriplet}), 
one arrives at the conclusion that, in the isosinglet case, the graphs of Fig.~\ref{3d} do not
generate extra contributions proportional to $g_{\mu\nu}$ either while, in the isotriplet case, such an extra term reads:
\be
8ig_{\mu\nu}\frac{g_Se^2}{\sqrt{2}}
\frac{\xi}{\beta^2}\frac{m^2}{(4\pi)^2}(\Lambda+1).
\label{defect}
\ee
In addition to the graphs depicted in Fig.~\ref{3d}, there are extra contributions due to the presence of new contact vertices 
(\ref{single}) and (\ref{double}) (see Fig.~\ref{2exd}). The single--photon vertex (\ref{single}) generates the contribution 
(see Fig.~\ref{2exd}(d))
\begin{eqnarray}
\nonumber
W_{\mu\nu}^{(1)}&=&g_Se^2\frac{\lambda_2}{\beta^2}i\epsilon_{3kn}\tau^{\alpha}_{k\beta}
\varGamma^{\beta}_{\lambda}\tau^{\lambda}_{n\rho}Q^{\rho}_{\alpha}\\
 & &\times \int\frac{d^4k}{(2\pi)^4}\frac{d^4q}{(2\pi)^4}
S(k)S(k-P)S(q)S(q-q_2)(2q-2k+q_1)_{\mu}(2q-q_2)_{\nu}\\
\nonumber  & & \qquad \qquad \qquad \qquad + (1\leftrightarrow 2, \mu \leftrightarrow \nu),
\end{eqnarray}
where the matrix structure of $\varGamma^{\beta}_{\lambda}$ is given
by either Eq.~(\ref{singlet}) or by Eq.~(\ref{triplet}). The
double--photon vertex (\ref{double}) gives rise to (see Fig.~\ref{2exd}(e)) 
\be
W_{\mu\nu}^{(2)}=2g_{\mu\nu}g_Se^2\frac{\lambda_2}{\beta^2}\varGamma^{\beta}_{\lambda}
(\vt^{\alpha}_{\beta}\vt^{\lambda}_{\rho}-
\tau^{\alpha}_{3\beta}\tau^{\lambda}_{3\rho})
\int\frac{d^4k}{(2\pi)^4}\frac{d^4q}{(2\pi)^4}S(k)S(k-P)S(q).  
\ee
Explicit calculations yield, for real photons, 
\be
W_{\mu\nu}^{(1)}+W_{\mu\nu}^{(2)}=0, 
\ee 
in the isosinglet case, and 
\be
W_{\mu\nu}^{(1)}+W_{\mu\nu}^{(2)}=-8ig_{\mu\nu}\frac{g_Se^2}{\sqrt{2}}\frac{\xi}{\beta^2}\frac{m^2}{(4\pi)^2}(\Lambda+1), 
\ee 
in the isotriplet case, so that this divergent term
cancels against that given by Eq.~(\ref{defect}). As a result, we
conclude that Eq. (\ref{Gexact}) holds also if exchange currents are
included.

\end{document}